\documentclass[aps,prl,showpacs, twocolumn,amsfonts,amssymb,amsmath]{revtex4}
\usepackage{graphicx} 
\usepackage{array}
\usepackage{SIunits}
\usepackage{color}
\newcommand{\OmegaO}[0]{\Omega_{\mathrm{0}}}
\newcommand{\omegaO}[0]{\omega_{\mathrm{0}}}
\newcommand{\omegaI}[0]{\omega_{\mathrm{1}}}
\newcommand{\OmegaI}[0]{\Omega_{\mathrm{1}}}
\newcommand{\ketO}[0]{\left|0\right>}
\newcommand{\ketI}[0]{\left|1\right>}
\newcommand{\PrYSO}[0]{Pr$^{3+}$:Y$_2$SiO$_5\,$}

\begin{document}

\title{Experimental quantum state tomography of a solid state qubit}

\author{L.~Rippe, B.~Julsgaard\footnote{present address: Research Center COM, DTU, DK-2800, Lyngby, Denmark}, A.~Walther, Yan Ying, S.~Kr\"{o}ll}
\affiliation{Department of Physics, Lund Institute of Technology,
  P.O.~Box 118, SE-22100 Lund, Sweden}

\date{\today }

\begin{abstract}
Full quantum state tomography is used to characterize the state of an ensemble based qubit implemented through two hyperfine levels in Pr$^{3+}$ ions, doped into a Y$_2$SiO$_5\,$ crystal. We experimentally verify that single-qubit rotation errors due to inhomogeneities of the ensemble can be suppressed using the Roos-M{\o}lmer dark state scheme \cite{Roos2004}. Fidelities above $>90\%$, presumably limited by excited state decoherence, were achieved. Although not explicitly taken care of in the Roos-M{\o}lmer scheme, it appears that also decoherence due to inhomogeneous broadening on the hyperfine transition is largely suppressed.
\end{abstract}


\maketitle

A large variety of systems are presently investigated in order to find out whether they can be used as hardware for quantum computers. The present work is carried out on a solid state based system, rare earth ions doped into inorganic crystals. As in several other solid state systems the qubits are encoded in nuclear spin states, which for rare earths can have coherence times of seconds and where much longer coherence times are predicted \cite{Longdell2006}. For being a solid state system the rare earth ions are unusual because their optical transitions can have coherence times as long as several ms \cite{Equall1994,Sun2002}. Quantum state tomography have previously been carried out to characterize the fidelity by which superpositions on an optical transition can be manipulated \cite{Longdell2004}. However, since coherence times for the hyperfine states are several orders of magnitude longer, it is highly relevant to also investigate the fidelity of arbitrary qubit rotations using hyperfine qubits. Multi-qubit gate operations can readily be implemented in the system, because optical excitation of an ion will induce frequency shifts $>$100 MHz ($>$10$^4$ line widths) of the optical transitions of nearby ions \cite{Ohlsson2002}. The large frequency shift of the optical transition makes it possible to entangle two nearby ions using operations with a duration of just a few ns \cite{Wesenberg2007}. A scalable implementation of the rare earth ion scheme can e.g. be achieved using a short lifetime readout ion, acting as a state sensitive probe for the local environment \cite{Wesenberg2007} in a manner similar to how the electronic spin of an NV center can probe the nuclear spin states of surrounding C$^{13}$ ions \cite{Childress2006}. However, because of the hour-long lifetimes of the rare earth spin states \cite{Koenz2003,Ohlsson2003}, it is possible to also create qubits consisting of an ensemble of ions, all in a specific quantum state. Each such qubit can be selectively manipulated by optical pulses \cite{Ohlsson2002,Rippe2005,Seze2005}. These ensemble qubits, which give strong readout signals, can be used to investigate general properties of the system. In this work ensemble qubits are used to experimentally carry out arbitrary rotations on the qubit Bloch sphere and the results are characterized by full quantum state tomography.\\
\begin{figure}[h]
		\includegraphics[width=8.5cm,height=3.3cm]{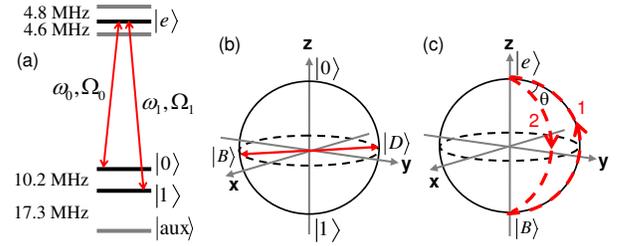}
	\caption{(color online) a) energy level diagram, b) qubit Bloch sphere, dark, $\left|D\right>$, and bright states, $\left|B\right>$, have been indicated, c) indicating the state transfer along paths 1 and 2 yielding an e$^{i\theta}$ phase shift of the bright state}
	\label{fig.1}
\end{figure}
The relevant part of the \PrYSO energy level diagram is shown in Fig.~1. The qubit states $\left|0\right>$\ or $\left|1\right>$ are represented by two of the three ground state hyperfine levels and the qubits can also be optically excited to the $\left|\mathrm{e}\right>$ state, which has a lifetime of 164  {\micro}s. Even if all ions in a specific qubit can be prepared in the $\left|0\right>$ or $\left|1\right>$ states using optical pumping, the ensemble approach brings additional problems because it has to be ascertained that all ions in a qubit have the same wave function. This is complicated by the facts that different ions within the qubit will; 1) have slightly different optical transition frequencies, leading to both different response to excitation pulses because some ions will be slightly off resonance and dephasing on the optical transition, 2) have slight differences in hyperfine transition frequencies, leading to different response to excitation pulses as well as dephasing on the qubit transition, 3) experience different optical field strength and have different Rabi frequencies, due to the spatial profile of the beam, which could result in different ions experiencing different pulse areas, finally, 4) when implementing, e.g., two-qubit gates the interaction between an ion in the first qubit and the nearby ion in the second qubit will be different for different instances. However, the scheme for two-qubit gates is designed such that it compensates for the inhomogeneity of the ion-ion interaction \cite{Ohlsson2002} and complications 1 and 3 have been solved by employing the techniques described in \cite{Roos2004}. Inhomogeneities in optical transition frequency and differences in Rabi frequency are compensated for by using complex hyperbolic secant pulses (abbreviated as \emph{sechyp pulses} through the rest of the text), which efficiently transfers states on the Bloch sphere from one pole to the other, provided the Rabi frequencies of the individual ions are above a certain lower limit. This was experimentally verified in Ref \cite{Rippe2005} and traces b) and c) in Fig.~2 also shows a transfer from the $\left|0\right>$ state to the $\left|1\right>$ state with about 96\% transfer efficiency. However, to demonstrate arbitrary operations on the qubit Bloch sphere, which is the main objective here, it is also necessary to address complication 2), the inhomogeneous broadening on the qubit transition. Since the \emph{sechyp pulses} only compensate for detunings in the resonance frequency and differences in Rabi frequency for pole-to-pole transfers on the Bloch sphere, Roos-M{\o}lmer \cite{Roos2004} introduced a basis change such that operations from an arbitrary point on the qubit Bloch sphere (Fig.~1b) could be implemented as pole-to-pole transfers in the new base. The concept is schematically pictured in Fig.~1 and briefly described below. Two fields with optical frequencies $\omegaO$ and $\omegaI$ and Rabi frequencies $\OmegaO$ and $\OmegaI$ are driving the $\left|0\right> - \left|e\right>$ and $\left|1\right> - \left|e\right>$ transitions, respectively. Adjusting the field amplitudes such that $\OmegaO = \OmegaI$ creates one bright state, $\left|B\right>$, and one dark state, $\left|D\right>$ 
\begin{equation}
\left|B\right>=\left|0\right> - e^{-i\phi}\left|1\right>, \:\left|D\right>=\left|0\right> + e^{-i\phi}\left|1\right>
\label{dark_def}
\end{equation}
depicted in the qubit Bloch sphere in Fig.~1b. $\phi$ is the relative phase difference between the two fields. The dark state wave function is not changed by the driving fields, however, the interaction between the driving field and the ions will drive the bright state along , e.g., path 1, on the $\left|B\right> - \left|e\right>$ Bloch sphere (Fig.~1c). If the fields $\OmegaO$ and $\OmegaI$ are \emph{sechyp pulses}, they will compensate for detunings on the optical transition frequency for the $\left|B\right> \rightarrow \left|e\right>$ transfer. If now the $\left|e\right> \rightarrow \left|B\right>$ transfer is carried out along a different path on the Bloch sphere, path 2, separated an angle $\theta$ from path 1, the bright state will have undergone the operation $\left|B\right> = e^{i\theta}\left|B\right>$. In the qubit basis ($\left|0\right>$, $\left|1\right>$), this is equivalent to the operation
\begin{equation}
U=e^{i\frac{\theta}{2}} \left( \begin{array}{cc}
\cos\frac{\theta}{2} & ie^{i\phi}\sin\frac{\theta}{2} \\
ie^{-i\phi}\sin\frac{\theta}{2} & \cos\frac{\theta}{2}
\end{array} \right) \label{arb_matrix}.
\end{equation}
Thus we are able to carry out unitary rotations about any axis on the equator of the Bloch sphere representing the qubit basis. The angles $\theta$ and $\phi$ can be set explicitly in the experimental environment, giving us high accuracy in creating the qubit rotation.\\
A Coherent 699-21 dye laser frequency stabilized against a spectral hole in a \PrYSO crystal yielding a coherence time $>$100 {\micro}s and a frequency drift $<$ 1 kHz/s was used for the experiments \cite{Julsgaard2006}. The light passed twice through an 200 MHz acousto-optic modulator such that light pulses with arbitrarily chosen phase, amplitude and frequency patterns could be created without any spatial displacement of the output beam. These pulses were then sent through a 350 MHz AOM driven by two rf frequencies separated by 10.2 MHz, which is the splitting between the qubit levels. In this way the first AOM produced the \emph{sechyp pulses} and the second AOM distributed this at frequencies $\omegaO$ and $\omegaI$, ascertaining that the pulses at the two frequencies had identical frequency chirp and amplitude variations (but different overall amplitude, phase and center frequency). After the AOM the light was passed through a single-mode fiber, to clean up the spatial mode. A few  percent of the light was split off after the fiber and used as a reference. The rest of the light (about 50 mW) was focused onto a 0.5 mm thick Y$_{2}$SiO$_{5}$ crystal where 0.05\% of the Y ions had been substituted by Pr$^{3+}$ to a 1/e$^2$ spot diameter of $\sim$100 {\micro}m, yielding a Rabi frequency of maximum 2 MHz for the strongest transitions. The light transmitted through the crystal was imaged onto a 50 {\micro}m pinhole only transmitting light from the center of the laser spot in the sample. Within this region the intensity varied by less than 30\%. Reference and signal beams were detected by two Thorlabs PDB150A detectors and the signals from the two detectors were divided to reduce the effect of laser amplitude fluctuations. The two-color \emph{sechyp pulses} used in the experiment had a duration of 4.4 {\micro}s and a FWHM of 1.2 {\micro}s\\
\begin{figure}[h]
		\includegraphics[width=8.3cm,height=7.2cm]{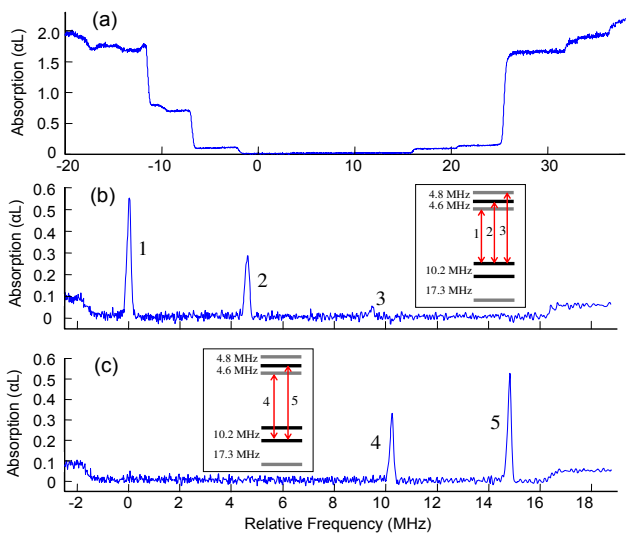}
	\caption{ (color online) a) Tailoring of the absorption profile, all ions within a 18 MHz spectral interval have been removed by optical pumping, b) a qubit is created by returning ions, which have their $\left|0\right> - \left|e\right>$ transition within 100 kHz from relative frequency 0, to the optically pumped region, c) The qubit has been transferred from the $\left|0\right>$ to the $\left|1\right>$ state. the transfer efficiency is about 96\%}
	\label{fig.2}
\end{figure}
The qubits are created as in Refs \cite{Rippe2005,Nilsson2004}. Fig.~2a shows a part of the inhomogeneously broadened Pr$^{3+}$ absorption line where all ions absorbing within an 18 MHz frequency interval have been transferred to other hyperfine states through optical pumping by repeatedly scanning the laser back and forth in frequency. Ions within a narrow frequency range are burnt back into the emptied frequency interval and placed in the $\left|0\right>$ state, creating a peak with an inhomogeneous width of about 170 kHz (Fig.~2b). Because of the upper state hyperfine splitting, the presence of ions in the $\left|0\right>$ state shows up in the absorption spectrum as three peaks separated by the upper state splittings, 4.6 MHz and 4.8 MHz. All spectra in Figs.~2~and~3 are recorded using the rapid chirp techniques developed in Refs \cite{Chang2005} and  \cite{Wolf1994}. Fig.~2c shows the absorption spectrum after first applying one \emph{sechyp pulse} at the $\omegaO$ frequency to the $\left|0\right> - \left|e\right>$ transition, bringing the entire population to the excited state, and then applying a second \emph{sechyp pulse} at the $\omegaI$ frequency on the $\left|1\right> - \left|e\right>$ transition, transferring the population to the $\left|1\right>$ state. The total transfer efficiency is about 96\%. The homogeneous dephasing time, $T_2$, for the optical transitions generally depends on the density of excited state ions \cite{Huang1989}. With a dopant concentration of 0.05 \% the density of excited ions for one excited, 170 kHz wide qubit is about $3\cdot10^{14}$/cm$^3$. From Ref.~\cite{Equall1995} this excited state density would give a excited state dephasing time of about 50 {\micro}s, which is consistent with our own photon echo measurements of the dephasing time when eciting one qubit. Simulating the state-to-state transfers with the \emph{sechyp} pulses used and a $T_2$ of 50 {\micro}s using a Bloch equation model, gives a maximum transfer efficiency of about 96\%. It is consequently reasonable to assume that excited state dephasing is the main limiting factor for the transfer efficiency in Fig.~2c.\\
We are now ready to use the two-color \emph{sechyp pulses}, employing the scheme outlined in Fig.~1 for performing arbitrary single qubit operations on qubits like the one shown in Fig.~2b. Starting from state $\left|0\right>$ (Fig.~2b) five different states on the Bloch sphere were prepared $\left|1\right>$, ($\left|0\right> + \left|1\right>$), ($\left|0\right> - \left|1\right>$), ($\left|0\right> + i\left|1\right>$) and ($\left|0\right> - i\left|1\right>$) (normalization factors have been omitted for notational simplicity). To characterize the created state we calculate the fidelity, \emph{F}, as \begin{math}F=\left<\Psi_{\mathrm{theory}}\right|\rho_{\mathrm{exp}}\left|\Psi_{\mathrm{theory}}\right>\end{math}, where $\Psi_{\mathrm{theory}}$ is the desired state and $\rho_{\mathrm{exp}}$ is given by \cite{Nielsen2000}
\begin{equation}
\rho_{\mathrm{exp}}=0.5\cdot[\mathrm{tr}(\rho)I+\mathrm{tr}(X\rho)X+\mathrm{tr}(Y\rho)Y+\mathrm{tr}(Z\rho)Z].
\label{rho_exp}
\end{equation}
I is the identity matrix, $X$, $Y$ and $Z$ are the Pauli matrices and tr($X\rho$), tr($Y\rho$) and tr($Z\rho$)) are the experimental results from measurements of the projection of the prepared qubit state on the $x$-, $y$- and $z$-axes and tr($\rho$) is set equal to unity (It is here assumed that decay to the $\left|\mathrm{aux}\right>$ state, Fig.~1a, can be neglected and indeed the branching ratio to this state is small \cite{Nilsson2004}.).
\begin{figure}[h]
		\includegraphics[width=7.9cm,height=7.9cm]{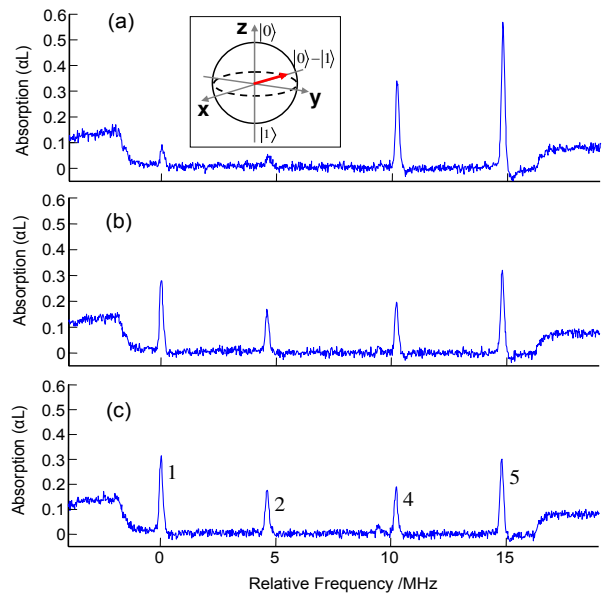}
	\caption{(color online) State tomography of the $\left|0\right> - \left|1\right>$ state. Traces a), b) and c) show the projection of the state on the $x$-, $y$, and $z$-axes, respectively. Numbers 1, 2, 4 and 5 in trace c), refer to the numbering of the transitions in Fig.~2. Further explanation is given in the text.}
	\label{fig.3}
\end{figure}
In Fig.~3 the arrow in the direction of the negative x-axis in the Bloch sphere schematically illustrates that the ($\left|0\right> - \left|1\right>$) state has been prepared. Traces a), b) and c) shows the results from the measurements of the projections of the ($\left|0\right> - \left|1\right>$) state on the $x$-, $y$- and $z$-axes. The only experimental measurement at hand is to measure the projection of a state on the $z$-axis, i.e., a frequency resolved absorption measurement telling the fraction of ions in the $\left|0\right>$ and the $\left|1\right>$ states. To measure the state projection on the $x$- and $y$-axes it is then necessary to carry out rotations on the Bloch sphere projecting these axes onto the $z$-axis. However, projecting the $x$-axis on the $z$-axis requires exactly the same pulses as creating the ($\left|0\right> - \left|1\right>$) state from the $\left|0\right>$ state and projecting the $y$-axis on the $z$-axis requires exactly the same pulses as creating the ($\left|0\right> -i\left|1\right>$) state from the $\left|0\right>$ state. Thus, there is now sufficient information to carry out the single-qubit operation and the state tomography. First a qubit is prepared as in Fig.~2a, then an operation on the $\left|B\right> - \left|e\right>$ Bloch sphere is done following the procedure in Fig.~1c where the angles $\phi$ and $\theta$ are chosen according to Eq.~\ref{arb_matrix} in order to carry out the desired operation. The qubit preparation and the qubit rotation is carried out three times. The first time the rotation is followed by a second rotation projecting the $x$-axis on the $z$-axis (followed by an absorption measurement, compare Fig.~3a), the second time the rotation is followed by a rotation projecting the $y$-axis on the $z$-axis, (compare Fig.~3b) and finally the rotation is followed by a projection of the $z$-axis on the $z$-axis by choosing $\theta$ in Fig.~1c equal to zero, bringing the bright state up and down along the same path. The $z$-axis projection can of course be measured by an absorption measurement directly, without any operations on the bright state, but actually carrying out the operation to project the $z$-axis onto the $z$-axis using $\theta = 0$, means that the $X$, $Y$ and $Z$ tomography measurements are all carried out in an equivalent manner. However, using only the pulses above the fidelity would be very low. The action of the two-color \emph{sechyp pulses} are to first promote the bright state to the excited state. In the general case the ions will then be in a superposition of the dark state and the excited state. Because of the inhomogeneous broadening of the optical transition, different qubit ions will acquire different phase factors while being in this superposition. However, a phase factor can be considered as global, and thus disregarded, if it appears in front of all qubit states. After the excited state has been returned to the bright state along path 2 in Fig.~1c , the detuning dependent phase factor accumulated in the excited state can now also be accumulated on the dark state. By sending in two new two-color \emph{sechyp pulses}, identical with the previous pair except that, first, the phase $\phi$ (see Eq.~\ref{dark_def}) now is increased by an amount $\pi$ (this means the new pulses will act on the dark state), second, $\theta$ is set equal to zero, which means the dark state is taken up and down on the $\left|D\right> - \left|e\right>$ Bloch sphere along the same route. Consequently U in Eq.~\ref{arb_matrix} will (apart from an overall phase factor) for this second \emph{sechyp pulse} pair, be an identity operation. However, the dark state part of the wave function will still have acquired the excited state phase factor due to inhomogeneous broadening. Dephasing due to inhomogeneous broadening on the optical transition is now eliminated. A further discussion of this issue can be found in \cite{Roos2004}. Thus, each single-qubit rotation as well as each tomography operation consists of four two-color \emph{sechyp pulses}. Two on the bright state and two on the dark state. In the present experiment each trace in Fig.~3a-c is the result after applying eight two-color \emph{sechyp pulses}, where each such pulse is 4.4 {\micro}s long. the total sequence for each trace then is 35.2 {\micro}s. The fidelities obtained are given in Table I.
\begin{table}[h]
	\centering
		\begin{tabular} {c|c c c c c c}
		  \hline  & $\ketO$ & $\ketI$ & $\ketO\! +\! \ketI$ & $\ketO\! -\! \ketI$ & $\ketO\! +\! i\ketI$ & $\ketO\! -\! i\ketI$\\
		   \hline F$_{\mathrm{QR}}$ & - & 0.96(2) & 0.93(1) & 0.93(1) & 0.92(2) & 0.91(2)\\
		   F$_{\mathrm{QR+QST}}$ & 1.02(2) & 0.92(3) & 0.87(1) & 0.87(2) & 0.85(4) & 0.84(4)\\	
		   \hline
		\end{tabular}
	\caption{ Fidelities for the single qubit rotation, F$_{\mathrm{QR}}$, and single qubit rotation + quantum state tomography, F$_{\mathrm{QR+QST}}$ for five different states. A quantum state tomography for the starting state, $\left|0\right>$, gives unity fidelity (within one standard deviation).} 
	\label{tab:1}
\end{table}
Two different fidelities are given F$_{\mathrm{QR+QST}}$ is the fidelity calculated according to Eq.~\ref{rho_exp}, where indices, QR and QST stand for Qubit Rotation and Quantum State Tomography, respectively. However, since the QR+QST operation just is two consecutive QR operations, it is reasonable to state the fidelity for a single qubit rotation, F$_{\mathrm{QR}}$, as ($F_{\mathrm{QR+QST}})^{1/2}$. Thus the fidelities for the single qubit operations are estimated to lie between 0.9 and 0.96, giving an average single operation fidelity of 0.93\\
The fidelities obtained are remarkably good considering the dephasing times of the system. The average time spent in the upper state during the QR+QST operation is $t_u$ = 8.8 {\micro}s. $e^{-t_u/T_2}$ = 0.84, which, assuming $\rho=0.84\rho(\mathrm{prepared}\:\mathrm{state})+0.16\rho(\mathrm{mixed}\:\mathrm{state})$, would give a best case fidelity of 0.92. Optically detected Free Induction Decay (FID) at the qubit transition, was used to determine the dephasing time on the hyperfine (hf) qubit transition, $T_2(\mathrm{hf})$. Using a two-color \emph{sechyp pulse} the qubit was put in a superposition state. The coherence on the qubit transition was probed by a delayed optical probe pulse. The 10.2 MHz beat signal due to the coupling between the qubit coherence and the transmitted optical pulse was detected. The strength of the beat signal as function of probe pulse delay measures the dephasing on qubit transition. The complete QR+QST sequence is 35.2 {\micro}s. From the FID measurement only about 20\% of the qubit coherence remained after 35 {\micro}s. The decoherence is reversible and caused by the inhomogeneous broadening of the qubit transition. In view of the fact that only about 20\% of the the qubit coherence remains after 35 {\micro}s, the fidelities obtained may seem remarkable. Possibly, this could be explained by the results in Ref, \cite{Tordrup2007}, where it has been shown that dynamical Stark shift occurring during qubit rotation can suppress errors due to inhomogeneous shifts of the qubit levels by as much as a factor of 10. It could also be a Zeno effect, which has been encountered previously in similar systems, such as NV centers \cite{Wrachtrup2006}. Still it would definitely be interesting to further investigate the effect of the inhomogeneous broadening on the QR fidelity.\\
The single qubit rotation fidelities could be improved; 1) by using pulses of shorter duration developed by optimal control theory \cite{Wesenberg2004b,Sporl2007}, 2) by using the single instance scheme \cite{Wesenberg2007} which eliminates errors due to inhomogeneous broadening on the hyperfine and optical transitions, or 3) by instead of the Pr ion use the Eu ion, where the upper state dephasing time is an order of magnitude longer. Assuming the fidelities in this work are limited by $T_2$ and $T_2(\mathrm{hf})$, one or several of these changes should enable fidelities above 0.99. To get fidelities significantly beyond this value, harder focusing of the light beam to increase the Rabi frequencies and/or dark state schemes not populating the excited state \cite{Goto2006,Goto2007}, would probably need to be used.\\
\emph{This work was supported by the European Commission through the ESQUIRE project and the integrated project QAP under the IST directorate, by the Knut and Alice Wallenberg Foundation, and the Swedish Research Council. B. Julsgaard was partly supported by the Carlsberg Foundation.}
\bibliographystyle{prsty}
\bibliography{../bibtex/Ref_lib}

\begin{thebibliography}{10}

\bibitem{Roos2004}
I. Roos and K. {M\o lmer}, Phys. Rev. A {\bf 69},  022321  (2004).

\bibitem{Longdell2006}
J.~J. Longdell, A.~L. Alexander, and M.~J. Sellars, Phys. Rev. B {\bf 74},
  195101  (2006).

\bibitem{Equall1994}
R.~W. Equall, Y. Sun, R.~L. Cone, and R.~M. Macfarlane, Phys. Rev. Lett. {\bf
  72},  2179  (1994).

\bibitem{Sun2002}
Y. Sun {\it et~al.}, J.Lumin. {\bf 98},  281  (2002).

\bibitem{Longdell2004}
J.~J. Longdell and M.~J. Sellars, Phys. Rev. A {\bf 69},  032307  (2004).

\bibitem{Ohlsson2002}
N. Ohlsson, R.~K. Mohan, and S. {Kr\"oll}, Opt. Commun. {\bf 201},  71  (2002).

\bibitem{Wesenberg2007}
J.~H. Wesenberg, K. Molmer, L. Rippe, and S. Kroll, Physical Review A {\bf 75},
   012304  (2007).

\bibitem{Childress2006}
L. Childress {\it et~al.}, Science {\bf 314},  281  (2006).

\bibitem{Koenz2003}
F. {K\"onz} {\it et~al.}, Phys. Rev. B {\bf 68},  085109  (2003).

\bibitem{Ohlsson2003}
N. Ohlsson, M. Nilsson, and S. {Kr\"oll}, Phys. Rev. A {\bf 68},  063812
  (2003).

\bibitem{Rippe2005}
L. Rippe {\it et~al.}, Phys. Rev. A {\bf 71},  062328  (2005).

\bibitem{Seze2005}
F. de~Seze {\it et~al.}, European Physical Journal D {\bf 33},  343  (2005).

\bibitem{Julsgaard2006}
L. Rippe, B. Julsgaard, A. Walther, and S. Kr\"{o}ll, submitted for publication
   (2007), quant-ph/0611056.

\bibitem{Nilsson2004}
M. Nilsson {\it et~al.}, Phys. Rev. B {\bf 70},  214116  (2004).

\bibitem{Chang2005}
T. Chang {\it et~al.}, Opt. Lett. {\bf 30},  1129  (2005).

\bibitem{Wolf1994}
F. Wolf, J. Phys. D {\bf 27},  1774  (1994).

\bibitem{Huang1989}
J. Huang, J.~M. Zhang, A. Lezama, and T.~W. Mossberg, Phys. Rev. Lett. {\bf
  63},  78  (1989).

\bibitem{Equall1995}
R.~W. Equall, R.~L. Cone, and R.~M. Macfarlane, Phys. Rev. B {\bf 52},  3963
  (1995).

\bibitem{Nielsen2000}
M.~A. Nielsen and I.~L. Chuang, {\em Quantum computation and quantum
  information} (Cambridge University Press, United Kingdom, 2000), eq. 8.148,
  Chapter 8.4.2.

\bibitem{Tordrup2007}
K. Tordrup and K. M{\o}lmer, Physical Review A {\bf 75},  022316  (2007).

\bibitem{Wrachtrup2006}
J. Wratchrup and F. Jelezko, J. Phys.: Condens. Matter {\bf 18},  807  (2006).

\bibitem{Wesenberg2004b}
J.~H. Wesenberg, Phys. Rev. A {\bf 69},  042323  (2004).

\bibitem{Sporl2007}
A. Sporl {\it et~al.}, Physical Review A {\bf 75},  012302  (2007).

\bibitem{Goto2006}
H. Goto and K. Ichimura, Physical Review A {\bf 74},  053410  (2006).

\bibitem{Goto2007}
H. Goto and K. Ichimura, Physical Review A {\bf 75},  033404  (2007).

\end{thebibliography}
\end{document}